\documentstyle[prl,aps,amssymb,twocolumn]{revtex}
\begin{document}
\draft
\title{Is Einstein's equivalence principle valid for a quantum particle?}
\author{Andrzej Herdegen\cite{AH}}
\address{Physics Department, University College Cork, Ireland\\{\rm and}\\
Institute of Physics, Jagiellonian University, ul.\ Reymonta 4, 30-059
Krak\'ow, Poland}
\author{Jaros{\l}aw Wawrzycki\cite{JW}}
\address{Institute of Physics, Jagiellonian University, ul.\ Reymonta 4,
30-059 Krak\'ow, Poland}
\maketitle
\begin{abstract}
Einstein's equivalence principle in classical physics is a rule stating
that the effect of gravitation is locally equivalent to the acceleration
of an observer. The principle determines the motion of test particles
uniquely (modulo very broad general assumptions). We show that the same
principle applied to a quantum particle described by a wave function on a
Newtonian gravitational background determines its motion with a similar
degree of uniqueness.
\end{abstract}

In this note we address one of the conceptual issues arising from the
efforts to reconcile quantum theory with gravitation, the question of the
status of the equivalence principle for quantum matter. In classical
physics the Einstein equivalence principle is a rule making one half of
the universal interdependence of geometry and matter, namely the
influence of geometry on matter, more specific. It states that the effect
of gravitation is locally indistinguishable from the effects arising from
the acceleration of the observer \cite{W}. Put differently, gravitational
effects may be locally ``transformed away'' by an appropriate choice of
the reference system. This is the principle used by Einstein himself;
some authors call it ``strong equivalence principle'' \cite{H}. We adopt
this version of the equivalence principle in this note as we believe that
it hits the heart of the matter, other ``equivalence principles'' (cf.\
below) being more accidental or secondary.

Within Einstein's gravitation theory one shows that the above equivalence
principle implies (modulo some natural assumptions on the general nature
of the equations of motion) that all test particles placed in this
spacetime move along geodesics. This fact is often expressed in one of
two ways: (1) that the motion of the particle is mass-independent, or (2)
that the inertial mass of the particle is equal to its gravitational
mass. These two statements are sometimes used interchangeably as ``weak
equivalence principle'' in the literature \cite{H}. This use of
terminology is rather confusing, as the two statements are logically
independent. They happen to coincide in the context of classical general
relativity, but may diverge in another setting. This is, as we shall see
below, what happens in the quantum case. The quantum dynamics of a test
particle following from the Einstein (``strong'') equivalence principle
is uniquely determined by one mass parameter. Thus the dynamics is
mass-dependent ((1) not true), but there is no room for independent
inertial and gravitational masses (in other words: the masses are equal,
(2) true ).

We turn now to this case. In the literature various opinions on the
status of the equivalence principle in the quantum world are expressed
\cite{art0}, and various, sometimes rather far removed from the original
geometrical notion of equivalence, ideas are proposed \cite{art1} (but
see also the final discussion). We think, however, that the extension of
the Einstein equivalence principle in the form stated above to the
quantum case experiences no logical difficulty, at least in the setting
in which it has often been considered. We feel, therefore, that it may be
of interest to see the simplicity of its action in this setting. The
setting referred to consists of a structureless particle described by a
wave function on a gravitational background of the nonrelativistic
spacetime (we use this term reluctantly: it is deeply rooted in the
physicists' jargon, but misleading; see below). This setting has been
adopted by several authors addressing the issue of covariance or
equivalence \cite{art1,art1a,art1b}. Within the path-integral formalism
conclusions similar to ours were reached earlier for the Feynamn
propagator of a structureless particle by De Bi\`{e}vre
\cite{art1b}. Our derivation, however, needs less assumptions, refers
directly to the wave function, and has the advantage of great simplicity,
both conceptual and technical (see also the closing discussion).

The reason for choosing the nonrelativistic rather than Einsteinian
spacetime is that we want to avoid the complications arising from
creation and annihilation of particles and their quantum
field-theoretical description, which has to replace (``first-quantized")
quantum mechanics in this case (there existing no consistent relativistic
quantum mechanics). The adopted setting is, however, nontrivial enough
and, in fact, contrary to the customary name, possesses a geometrical
structure (Newton-Cartan) interpretable in physical terms as a relativity
theory, but with Galilean rather than Lorentzian local inertial observers
\cite{art2}.

We can now state the main claim of this note. Consider a quantum particle
described by a wave function $\psi$ in a geometrical background with the
Newton-Cartan structure. Assume that the probability density of the
particle is a scalar field. Then the Einstein equivalence principle
determines the motion of the particle. This motion is not
mass-independent, but the inertial and gravitational masses are
necessarily equal (which is what one observes in experiment, see Ref.\
\cite{art3}; we shall return to the experimental aspect of the
equivalence principle in the concluding discussion). The choice of the
mass parameter is the only freedom of the equation. The equation itself,
when written in an appropriate coordinate system, is nothing else but the
usual Schr\"{o}dinger equation with the Newtonian potential term. We move
now on to the details.

We start by giving a brief account of the Newton-Cartan geometry. We
shall not discuss the underlying axioms and the logical structure of this
geometry, referring the reader to the existing literature \cite{art2},
but rather summarize the resulting structure in simple terms. The
Newton-Cartan geometry is defined on a four-dimensional differential
manifold. This manifold is equipped with an absolute time $t$ defining
the foliation of the spacetime by simultaneity hypersurfaces, a
positive-definite metric on each of these hypersurfaces, and a covariant
derivative (affine connection) compatible with these structures. However,
as a metric on a hypersurface is a \emph{form}, there is no unique way of
embedding it in the four dimensional manifold without additional
structures. This is how a gauge freedom in the choice of a
four-dimensional metric arises. Nevertheless, both the metric properties
on the hypersurfaces of constant time as well as the compatible
connection are unique (gauge-independent). With respect to the thus
defined connection the leaves of constant time are flat. The non-flatness
of the geometry reflects only the way in which the leaves fit together to
form the four-dimensional spacetime, and is encoded in one single scalar
field $\phi$. This field, however, is again non-uniquely determined by
the connection, being subject to a gauge freedom. All this sounds rather
more complicated than for a Lorentzian manifold of Einstein's theory of
gravity, but now a great simplification comes. In the Newton-Cartan
geometry there exists a class of privileged global coordinate systems,
the so-called Galilean coordinates. One of the coordinates in each of
these systems is always the time coordinate $t$ up to a translation by a
constant. The other will be denoted by $x^i$ ($i=1,2,3$). The space part
of the coordinate basis is a Cartesian system with respect to the metric:
$\left(\partial/\partial x^i\right)\cdot\left(\partial/\partial
x^j\right)=\delta_{ij}$. Moreover, vectors $\left(\partial/\partial
x^i\right)$ are parallel propagated by the connection, so the covariant
derivative of $\left(\partial/\partial t\right)$ gives the only
nontrivial characteristic of the connection, and must be expressible in
terms of $\phi$. In fact, with each choice of a Galilean system a natural
gauge of the field $\phi$ is chosen by the formulas: $\nabla_\mu
\left(\partial/\partial t\right)^\nu= \phi^\nu\nabla_\mu t$, where in the
coordinate basis the vector field $\phi^\nu$ is given by $\phi^0=0$,
$\phi^i=\left(\partial\phi/\partial x^i\right)$. This fixes the choice of
$\phi$ up to an addition of an arbitrary function of time. If $(t, x^i)$
is a Galilean system, then $(t', x'{}^i)$ is also a Galilean system if
and only if it is related to $(t, x^i)$ by a transformation of the form:
\begin{equation}\label{tr}
 t'=t+b\, ,\quad\vec{x}'=R\,\vec{x}+\vec{a}(t)\, ,
\end{equation}
where $R$ is an orthogonal transformation and $\vec{a}(t)$ is an
arbitrary time-dependent translation. Let us denote $(t,
\vec{x})\equiv X$, $(t',\vec{x}')\equiv X'$ and let us write the
transformation as $X'=r X$. The two fields $\phi$ and $\phi'$
correlated with the two systems are then related by the
transformation
$\phi'(X')=\phi(X)-\ddot{\vec{a}}(t)\cdot\vec{x}'+\mathrm{arbitrary\
function\ of\ time}$. We choose the function to be zero and write
the transformation law in the form
\begin{equation}\label{phi}
\phi'(X)=\phi(r^{-1}\! X) - \ddot{\vec{a}}(t-b)\cdot\vec{x}\, .
\end{equation}
The field $\phi$ is no longer a true scalar field, as with each of the
two systems a different gauge has been fixed.

The Newton-Cartan spacetime is flat if and only if there exists a
Galilean system in which $\phi$ is a function of time only (and may be
chosen equal to zero). The same holds then in all those Galilean systems
which are related by a transformation from the Galilean group
($\ddot{a}(t)=0$) to the first one (and, consequently, to each other).
These special Galilean systems are called global inertial systems. If the
spacetime is curved there are no global inertial systems, but, as is
evident from the transformation law (\ref{phi}) and the form of the
connection, for any chosen point of spacetime there exist Galilean
systems in which connection vanishes in this point. The restrictions of
these systems to a small neighborhood of this point are related to each
other by Galilean transformations and are called local inertial systems.
Both global (when they do exist) and local inertial systems have exactly
the same physical interpretation in terms of special observers as in
Einstein's theory.

The Newton-Cartan spacetime is the spacetime of the Newtonian world with
gravitation. The field equation of the form: ``the Ricci tensor of the
connection = 0" turns out to be identical in a Galilean system with the
Laplace equation for $\phi$, and the geodesic equation has in this system
the form of the Newton's second law for a particle in the gravitational
potential field $\phi$. In parallel with the Einstein theory of gravity
the geodesic law of motion of classical test particles may be obtained by
the application of the Einstein equivalence principle. And here again the
two minor equivalence principles are true in the classical case.

The existence of the global Galilean systems simplifies greatly the
investigation of the covariance of equations. To see whether an equation
has a geometrical, independent of the choice of coordinates, meaning, it
is sufficient to check whether it has the same form in all Galilean
systems. If it has, writing its coordinate independent form may pose some
technical difficulties, but is possible. In what follows we use the
Galilean systems only.

We are now prepared to place a quantum particle in this Newtonian
geometry. We assume that it is described by a wave function $\psi(X)$,
such that the correlated ``probability density"
$\rho(X)\equiv\bar{\psi}\psi(X)$ is a scalar field: $\rho'(X')=\rho(X)$.
We use this quantum mechanical language, but the argument is purely field
theoretic in spirit, and no a priori assumptions on the integration of
probability need to be made. The scalar transformation law of $\rho$ does
not determine the transformation law of $\psi$, as it says nothing about
the phase of $\psi$. Therefore, the problem to be solved is this: Can we
ascribe in each Galilean system a phase to the $\psi$ in such a way that
a consistent transformation law of the form
\begin{equation}\label{psi}
 \psi'(X)=e^{-i\theta(r,X)}\psi(r^{-1}\! X)
\end{equation}
would hold and $\psi$ would satisfy a form invariant equation in all
those systems?

Following standard assumptions we restrict considerations to the class of
linear equations of second order at most. Wishing to make use of the
equivalence principle we first have to answer this question for $\psi$ in
flat space, with the restriction of coordinate systems to inertial ones.
One could make use of the standard quantum-mechanical arguments, which
would produce the free particle Schr\"{o}dinger equation with the well
known transformation law of $\psi$ consistent in the quantum-mechanical
sense, as a projective unitary representation of the Galilean group.
However, we think that it is instructive to see how the same result
follows by purely geometrical arguments, without any use of a Hilbert
space. We sketch the argument briefly. The coordinate transformations are
now restricted to the Galilean group, $\vec{a}(t)=\vec{v}t+\vec{a}$ in
Eq.\ (\ref{tr}). We assume that the equation for $\psi$ has the following
form (the same in all inertial systems):
\[
 \left[
 a\,\partial_t^2+b_i\,\partial_i\partial_t+ c_{ik}\,\partial_i\partial_k+
 d\,\partial_t+f_i\,\partial_i+ g\,\right]\psi=0\, ,
\]
where $a(X)$, \ldots, $g(X)$ are the same functions in each coordinate
system, and $\psi$ transforms according to a law of the form (\ref{psi})
with $\theta$ to be determined. In the ``unprimed'' version of this
equation we substitute $\psi(X)=e^{i\theta(r,X')}\psi'(X')$ in accordance
with Eq.\ (\ref{psi}), express the differentiations in terms of the new
variables $X'$ and divide the resulting equation by the phase factor
function $e^{i\theta(r,X')}$. In the resulting equation the ratios of the
coefficient functions standing at the distinct differential operators
must be equal to the ratios of $a(X')$,
\ldots, $g(X')$. It is easy to see that the transformations of the
coefficient functions at the second order operators remain unaffected by
the phase function $\theta$. Considering first spacetime translations
$X'=X+Y$, $Y=(b,\vec{a})$, one finds, in particular, that the ratios of
$a(X)$, $b_i(X)$, $c_{ik}(X)$ must be equal to those of $a(X+Y)$,
$b_i(X+Y)$, $c_{ik}(X+Y)$ for all $X$ and $Y$. It follows that rescaling
the original equation by an appropriate factor function of $X$ one can
assume without loss of generality that $a$, $b_i$ and $c_{ik}$ are
constant. Consider now general Galilean transformations. For the first
two coefficients the covariance now demands that $a=\lambda a$,
$R\vec{b}+2a\vec{v}=\lambda \vec{b}$ (where $\lambda$ may be a function
of the transformation). These conditions can be satisfied for all $R$ and
$\vec{v}$ only if $a=0$ and $\vec{b}=0$. The transformation of the third
coefficient is now $R_{ij}R_{kl}c_{jl}=\lambda c_{ik}$. This may be
satisfied only if $c_{ik}=c\delta_{ik}$. Rescaling the equation by an
appropriate phase factor one can assume that $c\geq 0$. However, if $c=0$
the equation is at most of the first order, and then considerations
similar to ours show that $d=0$ and $\vec{f}=0$. This case is trivial.
Thus $c>0$ and $\lambda=1$. Now one writes the transformation of the
equation in full. From the invariance of the $\partial_t$-term one finds
that $d$ is a constant. The condition for $\vec{f}(X)$ takes the form
\[
 \vec{f}(X)=R^{-1}\left[
 \vec{f}(X')-d\vec{v}-2ic\vec{\partial}'\theta(r,X')\right]\, .
\]
Considering this condition for translations and rotations one finds that
${\mathrm{Re}}\,
\vec{f}(X)$ is an invariant vector field, thus ${\mathrm{Re}}\,
\vec{f}(X)=0$, and then $d=ik$, with real $k$. Applying
$\vec{\partial}\times$ to the imaginary part of the condition one finds
that also $\vec{\partial}\times{\mathrm{Im}}\,\vec{f}(X)$ is an invariant
field, so ${\mathrm{Im}}\,\vec{f}(X)=\vec{\partial}\, h(X)$. At this
point looking back to the equation we realize that what remains of this
term may be got rid off by a redefinition of the phase of $\psi$ (by
$-(i/2c)h(X)$). We can assume then that $\vec{f}(X)=0$ and find that
$\theta(r,X)=-(k/2c)\,\vec{v}\cdot\vec{x}+\tilde{\theta}(r,t)$. Finally,
the condition for $g(X)$ reads now
\[
 g(X)=g(X')-(k^2/4c)\vec{v}^2+ k\partial_{t'}\tilde{\theta}(r,t')\, .
\]
Applying $\vec{\partial}$ to this condition and considering translations
and rotations we find that $\vec{\partial}\, g$ vanishes, so
$g(X)=g(t)\equiv G'(t)$. Here again we realize that redefining the phase
of $\psi$ (by $(i/k)G(t)$) we remove the remaining freedom of the phase
in $\psi$ and the $g$ term from the equation. (If $k=0$ then $g$ is
invariant, so $g=\mathrm{const}$. We get the Helmholtz equation and
scalar transformation law for $\psi$. This, being not a dynamic equation,
we discard.) We finally get, with the standard notation of constants,
\[
 \left[ i\hbar\,\partial_t+\frac{\hbar^2}{2m}\,
 \vec{\partial}^2\right]\psi=0\, , \quad \theta
 (r,X)=-\frac{m}{\hbar}\vec{v}\cdot\vec{x}+\frac{m}{2\hbar}\vec{v}^2t\, ,
\]
which, of course, is the standard free particle theory.

Einstein's equivalence principle implies now that if in the flat space we
transform the Schr\"{o}dinger equation to all arbitrary Galilean systems
(noninertial), then we can identify all local modifications to the
equation which can appear in an arbitrary Galilean system in curved
spacetime. We assume the transformation law (\ref{psi}) and find that the
transformed equation differs from the Schr\"{o}dinger equation at most by
additional terms on the left-hand side of the form
$i\vec{\chi}\cdot\vec{\partial}\psi +
\Lambda\psi$, where $\vec{\chi}$ is real. In curved spacetime the
Einstein equivalence principle gives then the equation
\[
 \left[ i\hbar\,\partial_t+\frac{\hbar^2}{2m}\,
 \vec{\partial}^2+ i\vec{\chi}(X)\cdot\vec{\partial}+
 \Lambda(X)\right]\psi(X)=0\, ,
\]
where $\vec{\chi}(X)$ and $\Lambda(X)$ are now fields characterizing
geometry. We assume that these fields are determined locally by the
geometry. Assuming the transformation law of the form (\ref{psi}) and
demanding the covariance of the equation we find the condition
\[
 \vec{\chi}(X)=R^{-1}\left[\vec{\chi}'(X')- \hbar\dot{\vec{a}}(t)-
 \frac{\hbar^2}{m}\vec{\partial}'\theta(r,X')\right]\, .
\]
Applying $\vec{\partial}\times$ to this equation we find that
$\vec{\partial}\times\vec{\chi}$ is a vector field, in particular it is a
vector field with respect to rotations. But $\phi$, the only
characteristic of the geometry, is a scalar field with respect to
rotations, so there is no local way in which a $\vec{\chi}$ giving rise
to a nonzero vector field $\vec{\partial}\times\vec{\chi}$ can be formed
with it. Hence $\vec{\chi}(X)=\vec{\partial}\xi(X)$. We observe now, that
this longitudinal field may be absorbed into the phase of $\psi$ (with
the appropriate modification of $\Lambda$), so one can assume $\chi=0$.
The transformation condition then simplifies to
$\theta(r,X)=-(m/\hbar)\dot{\vec{a}}(t-b)\cdot\vec{x}
+\tilde{\theta}(r,t)$.  The covariance condition for the $\Lambda$ term
now takes the form
\[
 \Lambda(X)=\Lambda'(X')
 -m\ddot{\vec{a}}(t)\cdot{\vec{x}}'-(m/2)\dot{\vec{a}}^2(t)
 +\hbar\,\partial_{t'}\tilde{\theta}(r,t')\, .
\]
At this point let us look back once more to the flat space case and
assume that $X$ is an inertial system. Then $\Lambda(X)=0$ and we find
that the additional terms in the operator acting on $\psi'$ arising from
the non-inertiality of the system $X'$ are
$m\ddot{\vec{a}}(t'-b)\cdot\vec{x}'+ (m/2)\dot{\vec{a}}^2(t'-b)
-\hbar\,\partial_{t'}\tilde{\theta}(r,t')$. We learn two things. First,
the terms are real, so by the equivalence principle $\Lambda(X)$ is real
in general. Second, a change of coordinates produces definite terms up to
linear order in $\vec{x}$. The equivalence principle then implies that in
the general case it should be possible by a change of coordinates to
eliminate in the neighborhood of a given point $X_0$ terms independent
of, and linear in $\vec{x}-\vec{x}_0$. Put differently, it should be
possible to transform away the value and the first derivative
$\vec{\partial}$ of $\Lambda(X)$ at this point. Let us introduce
$\tilde{\Lambda}(X)$ by $\Lambda(X)= -m\, \phi(X)+ \tilde{\Lambda}(X)$.
Using Eq.\ (\ref{phi}) we find that the covariance condition now takes
the form
\[
 \tilde{\Lambda}(X)= \tilde{\Lambda}'(X')-\frac{m}{2}\dot{\vec{a}}^2(t)+
 \hbar\,\partial_{t'}\tilde{\theta}(r,t')\, ,
\]
which implies $\vec{\partial}'\tilde{\Lambda}'(X')=
R\,\vec{\partial}\tilde{\Lambda}(X)$. It is now clear that if
$\vec{\partial}\tilde{\Lambda}(X)\neq 0$ at some point then it cannot be
transformed away. Therefore $\tilde{\Lambda}(X)=\tilde{\Lambda}(t)$, and
may be removed by a change of phase of $\psi$. Thus the unique solution
for $\Lambda$ is $\Lambda=-m\phi$. We now see the geometrical meaning of
the condition that the first derivative $\vec{\partial}\Lambda$ may be
removed at a point by a change of coordinates: this derivative is
equivalent to the connection, so the meaning is exactly the same as in
the classical case. The covariance condition now simplifies to
$-(m/2)\dot{\vec{a}}^2(t) +\hbar\,\partial_{t'}\tilde{\theta}(r,t')=0$.
In this way we finally obtain the equation
\begin{equation}\label{schg}
 \left[ i\hbar\,\partial_t+\frac{\hbar^2}{2m}\,
 \vec{\partial}^2 -m\, \phi(t,\vec{x})\right]\psi(t,\vec{x})=0\, ,
\end{equation}
and the transformation exponents
\begin{equation}\label{th}
 \theta(r,X)=-\frac{m}{\hbar}\dot{\vec{a}}(t-b)\cdot\vec{x}
 +\frac{m}{2\hbar}\int_0^t \dot{\vec{a}}^2(\tau-b)\, d\tau \, .
\end{equation}

Until now we have considered the relation between two coordinate systems
only. Is the resulting structure, the equation (\ref{schg}) and the
transformation laws (\ref{phi}) and (\ref{th}), consistent with the
composition of transformations? That is, do we get the same result if we
choose to break the transformation $X\mapsto X'$ into two steps with an
intermediate system on the way: $X\mapsto X''\mapsto X'$? The answer is
that the two final gravitational potentials differ in general by a
time-dependent ($\vec{x}$-independent) additive term, while the two final
wave functions differ by a time-dependent phase factor. This, however,
poses no difficulty. The difference in the potentials is consistent with
the freedom in their definition, while a time dependent phase factor in
the wave function does not change the state vector (in
$L^2({\mathbb{R}}^3, d^3x)$) at any time, and in the equation induces
only another change of $\phi$ by an addition of a function of time. We
mention as an aside that one could begin the whole analysis by
classifying on group-theoretical basis all exponents $\theta(r,X)$
fulfilling this consistency condition. Such a classification has been
achieved by one of us (J.W.) and will be published elsewhere.

We have thus shown that Eq.\ (\ref{schg}) is uniquely determined by
Einstein's equivalence principle. In particular, we have shown that the
principle implies equality of inertial and gravitational masses. The
equation, of course, is standard, and has been discussed many times, but
the derivation of its geometrical uniqueness within standard quantum
mechanics is new. Within the path-integral formalism De Bi\`{e}vre
obtained earlier similar results for the Feynman propagator of a particle
in gravitational field. His derivation is based on an additional
geometric structures (the Bargmann bundle over spacetime and an
associated vector bundle). The propagator is assumed to have geometric
properties with respect to this structures. As the connection of this
formalism with the standard quantum mechanics is not explicit at this
stage it is not quite obvious what are the corresponding restrictions on
the wave function. However, they must amount at least to some
restrictions on the transformation properties of the wave function. On
the other hand in our derivation the phase of the wave function is
completely free at the start. The equivalence principle and the
adjustment of the phase yield both the dynamic equation and the
transformation law in an extremely simple way.

Within standard quantum mechanics Kucha\v{r}
\cite{art1a} has derived the equation (in general covariant form) by
canonical quantization of the geodesic motion. (Where in the process is
the mass independence lost? It is, of course, when after going over to
the Hamiltonian formalism, in which mass appears, the momentum looses any
memory of the mass upon replacement by $-i\hbar\vec{\partial}$.) However,
canonical quantization is a heuristic procedure (it is rather classical
mechanics, which is believed, in principle, to be derivable from the
quantum mechanics) and it is unable to decide the uniqueness question or
to clarify the intrinsic structure at play. On the other hand Duval and
K\"{u}nzle \cite{art1a} work from start with a wave function of a
particle. They show how the equation obtained by Kucha\v{r} may be
derived by the minimal coupling principle if the wave function is assumed
to have certain geometrical properties (is a section of a vector bundle
associated with the Bargmann bundle over spacetime). The geometric
structures introduced by them have been then adopted by De Bi\`{e}vre in
the paper mentioned above, and also by Christian \cite{art1a}, who makes
it a basis for a construction of a Newton-Cartan quantum field theory of
particles and gravitational field. While the structures introduced by
Duval and K\"{u}nzle illuminate the geometry of the general covariant
Schr\"{o}dinger equation, they incorporate assumptions on the
transformation properties of the wave function and on the form of the
equation which we derive here from scratch.

Another approach to the question of the validity of equivalence principle
in the quantum world has been proposed by L\"{a}mmerzahl
\cite{art1}. He gives arguments to the effect that Eq.\ (\ref{schg}) is
favored by a principle which he introduces and calls quantum equivalence
principle. This principle formulates a condition for a possibility of the
extraction of mass-independent characteristics from experimental results.
However, there is no obvious connection of this principle with Einstein's
geometrical idea and its compelling persuasiveness (in fact,
L\"{a}mmerzahl avoids the covariance question completely). We do believe
L\"{a}mmerzahl's results are important and interesting, but see their
role on the experimental side rather than as a theoretical paradigm. What
we mean, more precisely, is this. Einstein's principle is a local
principle. For a classical test particle, which is a local object, its
content translates itself rather directly into experimental predictions.
The quantum mechanical wave function, on the other hand, is a nonlocal
object, and there is no simple analogous translation - in general gravity
cannot be eliminated on any hypersurface of constant time.
L\"{a}mmerzahl's papers show how to extract experimental consequences of
Einstein's equivalence principle from experimental data. Having said
this, however, we also want to express disagreement with the opinion that
nonlocality of the wave function precludes operational meaning of
Einstein's principle. It may be not obvious how to reveal such meaning,
but we can see no fundamental obstacle in the way. Measurements are done
locally, which is the operational foundation of the Einstein's principle.

\end{document}